\documentstyle[aps,epsf,twocolumn]{revtex}
\newcommand{\bea}{\begin{eqnarray}}
\newcommand{\beq}{\begin{equation}}
\newcommand{\eea}{\end{eqnarray}}
\newcommand{\eeq}{\end{equation}}
\begin{document}
\title{Self-Consistent Harmonic Oscillator Model
and Tilted Rotation}
\author{W.D. Heiss$^{\star}$ and R.G. Nazmitdinov$^{\star,\star \star}$}
\address{$^{\star}$ Centre of Theoretical Physics and School of
Physics\\
University of the Witwatersrand, PO Wits 2050, Johannesburg, South
Africa\\
$^{\star \star}$ Bogoliubov Laboratory of Theoretical Physics,
Joint Institute for Nuclear Research, 141980 Dubna, Russia}
\maketitle
\today

\baselineskip 20pt minus.1pt

\begin{abstract}
The three dimensional harmonic oscillator model including a
cranking term is used for an energy variational calculation.
Energy minima are found under variation of the three oscillator
frequencies determining the shape of the system for given values of the
three components of the rotational vector which determines the
orientation of the angular momentum in the intrinsic frame.
Tilted rotations are established by numerical means for triaxial
nuclei. The onset of tilted rotations is related to an
instability of the mean field as verified by the random
phase approximation. Analytic expressions are derived for
the critical rotational frequencies associated with the mean
field instability.
\end{abstract}
\vspace{0.2in}

\vspace{0.2in}

\section{Introduction}

Numerous experimental observations of  well developed rotational bands
implying  $\Delta I=1$ sequences (see for a review \cite{Fr01}) raise
the question about mechanisms creating such
collective states. The conventional description within the
Cranking Shell Model (CSM) \cite{NR} suggests the $\Delta I=2$
sequence of rotational states. As is well known, in the CSM the axis of
rotation coincides with the principal axis of the density distribution
of the nucleus (principal axis cranking or PAC).
As a result, the CSM Hamiltonian adheres to
the $D_2$ spatial symmetry with respect to rotation
by the angle $\pi$ around the rotational axis,
for instance, the $x-$ axis.
Consequently, all rotational states can be classified by the quantum number called
signature $r=\exp (-i\pi\alpha)$ leading to selection
rules for the total angular momentum
$I=\alpha + 2n$, $n=0,\pm 1, \pm 2 \ldots $.
In particular, in even-even nuclei the lowest rotational (yrast) band
characterised by the positive signature quantum number
$r=1 \,(\alpha = 0)$ consists of even spins only.

It was suggested \cite{Fr93} that non-principal axis rotations
which invalidate the concept of signature symmetry is
responsible for rotational bands with $\Delta I=1$ sequences.
This idea has led to considerable experimental and theoretical efforts
during the past few years \cite{Fr01}. Related questions have been
addressed recently in connection with
effects of superfluidity in a trapped Bose-Einstein condensate (BEC)
\cite{St01}.

For a nucleus, the non-principal axis rotation is considered
stationary. For the BEC, a precession of the symmetry
axis is assumed for the condensate around the symmetry axis of the confining trap
with the precession frequency being fixed by the angular momentum
carried by the vortex line. A quantised vortex is aligned along the
symmetry axis of the condensate carrying one unit
$\hbar$  of angular momentum per particle.
Similarly, the precessional motion of the
principal axes in the nucleus creates vibrational (wobbling) excitations
above the yrast line \cite{BM75,JM,Mar}.
Notice that in this case the wobbling excitations are quantum mechanical
fluctuations of the angular momentum around a fixed principal axis of
the mean field generated by the CSM \cite{JM,Mar}. 
These excitation carry one unit
of angular momentum, similar to the vortex in the BEC, and
corresponding states may be associated with odd spin partners of
the gamma band in a rotating nucleus at low angular momenta \cite{MJ}.
It has been demonstrated \cite{Jan78} that at high rotational frequency
the wobbling excitations become particularly low. Similar to the transition
from spherical to deformed nuclei \cite{BM75} when quadrupole 
excitation energies tend toward zero, an analogous mechanism 
is expected to prevail in that for
vanishing wobbling excitations the nucleus undergoes a transition to tilted
rotation as the energetically favoured state.
The analysis of the link between tilted rotation and wobbling motion is one
major aspect of the present paper.

In fact, a planar tilted rotation
effecting the transition from the superfluid
to the normal  phase has been
discussed within the phenomenological two-phase model \cite{Ma78}.
The first microscopic attempt to describe a non-principal axis rotation
has been carried out in \cite{FB} using
the triaxial cranked oscillator model (TCO) with constant angular
frequency (see for a review of latest studies \cite{Fr01}).
However, the TCO calculations employ a fixed triaxial
deformation. The violation of the self-consistency between rotation and
evolution of shape parameters is a serious  drawback of this model.
In the present paper we investigate
the occurrence of tilted rotation using the self-consistent triaxial
cranked oscillator model. The model guarantees full
self-consistency at all rotational frequencies and all particle numbers.
A major outcome of the present paper is the result that, in even-even nuclei,
tilted rotations occur {\it if and only if the nucleus has a
triaxial shape in its ground state}.
The pairing interaction could have an effect upon the conclusions
drawn in the present paper.
However, its role is less important at high spins where
tilted rotations should be observed \cite{Fr01}. On a technical footing,
it involves additional variational parameters which complicates
the minimisation procedure. Therefore, we exclude in our model the pairing
interaction. The transparency of the model which provides a general
tendency of the phenomena and the succinct results compensate for this shortcoming.

In Sec.II we review the main features of our model. The 
numerical results are presented in Sec.III.
The results are discussed in Sec.IV
 and a short summary in Sec.V.
Technical details are deferred to Appendices.
First preliminary results have been reported in \cite{tilt,wobb}.

\section{The model}

The many-body Hamiltonian (Routhian) in the rotating frame is given by
\beq
\label{ham}
H=\sum_{i=1}^N(h_0(i)-\vec \Omega \cdot \vec l(i))
=H_0-{\vec{\Omega}}\cdot \vec{L}
\eeq
where the single particle triaxial harmonic oscillator Hamiltonian $h_0$ is
aligned along its principal axes and reads
\beq
h_0={1\over 2m}{\vec  p}^2+{m\over 2}
(\omega _x^2 x^2+\omega _y^2 y^2+\omega _z^2 z^2).
\eeq
The rotational vector $\vec \Omega $ of the cranking term has the
components $(\Omega _x,\Omega _y,\Omega _z)=\Omega (\sin \theta \cos \phi,
\sin \theta \sin \phi ,\cos \theta ).$

In Appendix A it is shown how this Hamiltonian is cast into the form
\begin{equation}
H=\sum_{j=1}^{N}\sum_{k=1}^{3} E_k (Q^{\dagger }_k Q_k+1/2)_j
\label{qham}
\end{equation}
where the normal mode energies $E_k$ are obtained from a cubic
equation and the normal mode operators $Q_k$ are linear
superpositions of the $p_i$ and $x_i$. In this way we have the
exact single-particle energies and wave functions at our disposal.
Note that, for $\Omega \ne 0$ and in particular for a tilted
rotation, the normal mode operators $Q_k$ are superpositions of
all three Cartesian components of the $p_i$ and $x_i$.
Furthermore, the eigenmode energies $E_k$ are real only for a
certain range of $\Omega \le \Omega _{{\rm max}}$ (see Appendix
B). We will restrict ourselves to this physical range of values
of the rotational frequency.

The knowledge of the eigenmodes and eigenfunctions enables us to calculate
energy contours of the total energy
\beq
E_{{\rm tot}}=
E_1\Sigma_1+E_2\Sigma_2+E_3\Sigma_3
\label{etot}
\eeq
in, say, the $\omega _x-\omega _y$ plane
with the particle number $N$ and the rotational vector $\vec \Omega $ as 
parameters. The configurations are determined by 
$\Sigma_k=\Sigma_j^N(n_k+1/2)_j$ where
the occupation numbers $n_k$ are the eigenvalues of $Q^{\dagger}_kQ_k$, they
take the values $0,1,2,\ldots $. For the PAC rotation around the
$x-$axis the notation
$\Sigma_1\equiv \Sigma_x$, $\Sigma_2\equiv\Sigma_+$,
$\Sigma_3\equiv \Sigma_-$ is used \cite{TA} (we chose
$\omega_x \geq \omega_y \geq \omega_z$, see also Appendix B).

The ground state is determined by filling
the single-particle levels from the bottom. We take care of the particle
spin only in obeying the Pauli principle which allows two particles in one
level; also we consider only one kind of nucleons, protons or neutrons.
It is clear that different sets of normal modes yield different sets of
occupation numbers. A pronounced shell structure can exist only 
for special sets of normal modes.
Correspondingly, all spherical shells will be strongly mixed in
our model, in particular the $\Delta N=2$ mixing is taken into account
exactly. In this way we go beyond the calculation of \cite{NR,Ros} where 
mixing between major oscillator shells is ignored.

The normal modes depend on the three components of the rotational vector
and on the harmonic oscillator frequencies. From our assumption
that the system adjusts itself under the influence of the rotation
by minimising $E_{{\rm tot}}$, a change of the magnitude or direction of
the rotation leads to a corresponding change of the effective mean field
potential which is given by the oscillator frequencies. In other words,
for a given rotational frequency, we must seek the minimum
of $E_{{\rm tot}}$ under variation of the oscillator frequencies.
The variation cannot be unrestricted as the confining potential
encloses a fixed number of particles, and assuming that
the particle density does not change we are led to a fixed volume
constraint
\beq
\omega _x\omega _y \omega _z={\omega _0}^3
=(41)^3/N.
\eeq
The volume constant is chosen in units where frequencies are given
in MeV and lengths in Fermi.
Introducing the Lagrange multiplier $\lambda$,
we solve the variational problem
\beq
\delta (\langle g|H|g\rangle -\lambda \omega _x\omega _y\omega _z)=0
\label{var}
\eeq
where $|g\rangle $ denotes the ground state as described above.

From Eq.(\ref{var}) we obtain, after differentiation with respect to the
oscillator frequencies and using Feynman's theorem \cite{Fey}
\begin{equation}
{d\over d\omega _k}\langle g|H|g\rangle =
\langle g|{dH\over d\omega _k}|g\rangle ,  
\label{feyn}
\end{equation}
the self-consistency condition
\begin{equation}
\omega _x^2\langle g|x^2|g\rangle=\omega _y^2\langle g|y^2|g\rangle
=\omega _z^2\langle g|z^2|g\rangle
\label{cond}
\end{equation}
which must be obeyed at the minimum of $E_{{\rm tot}}$.
Notice, that Eq.(\ref{cond}) becomes a general result
which includes as a particular case the cranking harmonic
oscillator with the PAC rotation around the $x$-axis \cite{TA}. 
The condition serves as an
acid  numerical test. Note that it
is fulfilled irrespective of the minimum being local or global.

For $\Omega=0$ we recover the well known magic numbers of the triaxial
harmonic oscillator for one type of particles (proton or
neutron) \cite{BM75} at $N=20,40,70,112,\ldots $ which
manifest themselves in our procedure by the spherical symmetric solutions
$\omega_x=\omega_y=\omega_z=\omega_0$. 
Between these magic numbers the well known
prolate or triaxial near prolate and oblate or triaxial near oblate
solutions are found, where a (near) prolate solution
($\omega_x\simeq \omega_y>\omega_z$) is obtained for a shell which is
less than half completed and a (near) oblate solution
($\omega_x\simeq \omega_y<\omega_z$) for a more than half completed shell.
In Fig.1 the values of $\langle x^2 \rangle,\,\langle y^2 \rangle $ and
$\langle z^2 \rangle $ are plotted versus $N$.
From the figure we see, that,
for $\Omega =0$, some shapes are axially symmetric while others are not.
Switching on the rotation $\Omega \neq 0$ a similar picture is obtained
as long as the rotational frequency is kept at a reasonably low value
($\Omega \le 1.0$ for $N\le 24$ and $\Omega \le 0.3$ for $N\ge 40$).
The corresponding minima of the total Routhian are a function of
$\vec \Omega$, i.e.~of
$\Omega,\theta $ and $\phi $. Our interest is focused upon the lowest
minimum under variation of $\theta $ and $\phi $ for fixed $\Omega $, i.e.
$\partial E_{{\rm tot}}/\partial \theta=\partial E_{{\rm tot}}/\partial
\phi=0$. If such minimum is found for $\theta \ne 0$ we refer to
this solution as to a tilted solution.

\section{Numerical Results}

As is well known \cite{TA}, the PAC rotation of prolate systems
leads to triaxial shapes. With increasing rotational speed the change of
the shape leads eventually, for a critical frequency
$\Omega _{{\rm crit}}^{(1)}$, to an oblate
shape with the rotational axis coinciding with the symmetry axis. Using the
minimising procedure of our model confirms this fact in principle. However,
our model allows an easy assessment as to whether the minimum is local or
global. It turns out that, for $N<56$, the minimum found
at $\Omega _{{\rm crit}}^{(1)}$ is no longer a global minimum,
since somewhere in the interval
$[0,\Omega _{{\rm crit}}^{(1)}]$ another lower lying minimum comes up.
As a typical example we consider the celebrated nucleus $^{20}$Ne
(since we take into account one kind of particle, this nucleus corresponds
to system with ten particles).  For $\Omega =0$ the global minimum occurs for
a prolate shape with the configuration
$(\Sigma _1,\Sigma _2,\Sigma _3)=(7,7,11)$.
With increasing $\Omega $ the minimum initially continues to be
global and the configuration remains constant while the nucleus becomes
triaxial. However, for $\Omega \approx 4.55$ another lower minimum
occurs with a completely
different, probably non-physical configuration
($(\Sigma _1,\Sigma _2,\Sigma _3)=(5,5,25)$). Only if the original
configuration is enforced, the oblate shape occurs at the larger value
$\Omega _{{\rm crit}}^{(1)}\approx 5.65$, but, being a local minimum,
it may have to be interpreted as an isomeric state.

\begin{figure}
\epsfxsize=2.2in
\centerline{
\epsffile{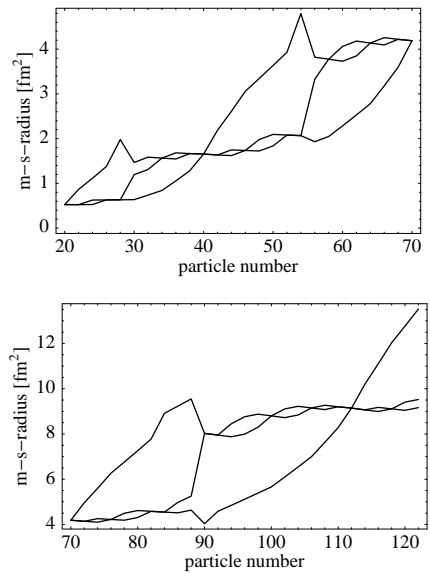}}
\vglue 0.15cm
\caption{ Mean square radii $\langle x^2\rangle,\,\langle y^2\rangle $ and
$\langle z^2\rangle $ versus particle number for minimal energy configurations
without rotation in the triaxial harmonic oscillator. The magic (spherical),
axially symmetric and triaxial configurations are clearly discernible.
}
\label{fig1}
\end{figure}

The observation, that the onset of an oblate shape is attained at a local
and not at a global minimum, prevails for all prolate nuclei with $N\le 56$.
When looking at Fig.1 there are the nuclei which have, for $\Omega =0$,
a triaxial shape close to the prolate shape, for instance $N=22,44,48,50$.
Initially when $\Omega $ is switched on, they behave just like the case
considered in the previous paragraph in that they reach eventually,
at $\Omega _{{\rm crit}}^{(1)}$, the oblate shape at a local minimum.
For a certain value $\Omega _{{\rm crit}}^{(2)}>\Omega _{{\rm crit}}^{(1)}$
the onset of tilted rotation arises. In other words, enforcing the initial
configuration and thus pertaining to the local minimum, the minimum is
now found at $\theta >0$ and possibly $\phi >0$ while the minimum
prevailing within the interval $[\Omega _{{\rm crit}}^{(1)},
\Omega _{{\rm crit}}^{(2)}]$ for $\theta =0$ has, at 
$\Omega =\Omega _{{\rm crit}}^{(2)}$, become unstable,
in fact it turns into a maximum in the $\theta -\phi $ plane.
In the following section we derive analytic expressions for
$\Omega _{{\rm crit}}^{(2)}$ and $\Omega _{{\rm crit}}^{(1)}$,
and provide an explanation why the tilted rotation
cannot occur when starting with a prolate nucleus irrespective
of the minimum being local or global. In this context we stress that
the model has of course its limitations and must be used with due 
discretion. The occurrence of new global minima for increasing values 
of $\Omega $ is an expected feature of the model which does not 
necessarily have a physical significance. For
light prolate nuclei, where the onset of oblate shapes occurs at values of
$\Omega $ so large that the minimum is not global, only a more sophisticated
approach could ascertain as to whether such a minimum is stable. 
Likewise, there is a temptation to associate the new global minimum 
with fission, but this also lies
beyond the scope of the present paper.

In contrast, for
all oblate (and heavy prolate nuclei) the minima at
$\Omega _{{\rm crit}}^{(1)}$ are global as the
critical values are sufficiently small. In these
cases we therefore view our findings as reliable and physically significant.
As is shown in the following section analytically, tilted
solutions can occur at $\Omega _{{\rm crit}}^{(2)}>\Omega _{{\rm crit}}^{(1)}$ 
and, in fact do occur, when starting from a nucleus
that is triaxial for $\Omega =0$. In turn, when starting from an oblate
nucleus, a tilted solution never occurs. Yet, the pattern is very similar
in both cases for $\Omega <\Omega _{{\rm crit}}^{(1)}$, in that
once the nucleus has become triaxial for $\Omega >0$, it attains the
oblate shape for $\Omega =\Omega _{{\rm crit}}^{(1)}$, where, for
the energy minimum, the rotational axis coincides with the symmetry axis.

\section{Bifurcation points}

The first subsection is devoted to the basics of our
approach with particular
emphasis on the extension of established knowledge.
In the following subsections the calculation
and the role of critical frequencies are discussed.

\subsection{Basics}

The two critical frequencies, $\Omega _{{\rm crit}}^{(1)}$ and
$\Omega _{{\rm crit}}^{(2)}$, are closely related to an
instability of the mean field. When $\Omega $ is sweeping beyond
the critical value, the mean field changes and may lead to a
symmetry breaking. The instability can be identified as a
bifurcation point \cite{Mar96} where new solutions that break
the symmetry emerge. As is demonstrated below, they are
related to collective vibrations carrying nonzero angular
momentum aligned along the symmetry axis. The bifurcation points
are defined by the vanishing vibrational frequencies in the
rotating frame. That occurs at a discrete set of cranking
frequencies given by 
\beq 
\Omega_{\rm crit}\equiv
\omega_{K}/K 
\label{mb}
\eeq 
where $K$ is the
angular momentum along the symmetry axis carried by a collective
vibration with the energy $\omega _K$ in the
laboratory frame.

We treat the collective excitations in the random phase approximation (RPA).
The variation in the one-body potential around the equilibrium shape
determines the effective quadrupole-quadrupole interaction \cite{Kis}.
As  a result, the total Hamiltonian can be presented as
\beq
H_{{\rm RPA}}=H_0-\vec \Omega \cdot \vec L
- {\kappa \over 2}\sum _{\mu =-2}^2 D_{\mu }^{\dagger} D_{\mu }
={\tilde H} - \vec \Omega \cdot \vec L
\label{hrpa}
\eeq
where the quadrupole operators $D_{\mu}=r^2Y_{2\mu}$ 
are expressed in terms of the double-stretched coordinates
$\bar q_i=\frac{\omega_i}{\omega_0} q_i$, $(q_i=x,y,z)$ 
(see also Appendix C).
Note that the effective interaction restores the rotational
invariance of the Hamiltonian $H_0$ in that now
$[{\tilde H},L_i]=0 \quad  (i=x,y,z)$.
Also note that the self-consistency conditions
Eq.(\ref{cond}) fix the quadrupole strength
$\kappa=\frac{4\pi}{5}\frac{m\omega_0^2}{<r^2>} $,
where $<r^2>=<\bar x^2+\bar y^2+\bar z^2>$.

We solve the RPA equation of motion for generalized coordinates
${\cal X}_{\lambda}$ and momenta ${\cal P}_{\lambda}$
(see for details \cite{KN}):
\bea
[H_{{\rm RPA}},{\cal X}_{\lambda}]&=&
-i\omega_{\lambda}{\cal P}_{\lambda},\quad
[H_{{\rm RPA}},{\cal P}_{\lambda}]=
i\omega_{\lambda}{\cal X}_{\lambda}, \\  \nonumber
[{\cal X}_{\lambda},{\cal P}_{\lambda}]&=&
i\delta_{\lambda, \lambda^\prime }.
\eea
Here, $\omega_{\lambda}$ is the RPA eigenfrequency in the rotating
frame and the associated phonon operator is
$O_{\lambda}=({\cal X}_{\lambda}-i{\cal P}_{\lambda})/\sqrt{2}$.
In contrast to the CRPA approach  \cite{JM,Mar,KN} 
developed for the PAC rotation,
the phonon is in the present model
a superposition of different signature phonons.
The degree of the mixture depends on the tilted angle:
the signature and $|K|$ are good quantum numbers,
respectively, for rotations coinciding with one of the principal
axes. 
The solution of the RPA equations leads to the secular
equation $F(\omega_{\lambda})=0$ with $F$ being a five by five determinant.
The roots are the RPA eigenfrequencies $\omega_{\lambda _i}$.
The non-zero solutions appear in pairs $\pm
\hbar \omega_{\lambda}$, we choose solutions with positive norm.
The function $F$ factors into positive and negative
signature parts for PAC rotations, since the signature
is then a good quantum number.  This simplifies
the calculations of the RPA modes and provides the analytical
solutions and the critical frequencies
$\Omega _{{\rm crit}}^{(1)}$ and $\Omega _{{\rm crit}}^{(2)}$, where
the minimal solution corresponds to the PAC rotation
around the symmetry axis being the $x$--axis.

Since at this transition point the minimal solution corresponds
to a one--dimensional rotation ($\vec \Omega\equiv (\Omega, 0,
0)$) around the symmetry axis, the angular momentum is a
good quantum number. The RPA states are characterised by the
projection of the angular momentum upon the symmetry axis ($x$--axis)
because
$[L_x,O_{\lambda}^{\dagger }]=\lambda O_{\lambda}^{\dagger }$
with $\lambda$ being the value of the angular momentum carried by
the phonons along the $x$--axis.
We thus obtain
\bea
[H_{{\rm RPA}},O_{\lambda}^{\dagger}]&=&[{\tilde H} -
\Omega L_x, O_{\lambda}^{\dagger }]=\nonumber\\
&&({\tilde \omega}_{\lambda}-
\lambda \Omega) O_{\lambda}^{\dagger } \equiv \omega_{\lambda}
O_{\lambda}^{\dagger }.
\label{man}
\eea
This equation implies that at the rotational frequency
$\Omega_{\it cr}={\tilde \omega}_{\lambda}/\lambda$ 
one of the RPA frequency vanishes.

\subsection{Critical frequencies for oblate rotation}
\subsubsection{Mean field}

Let us first turn to our numerical findings relating to oblate and/or
near oblate cases. In these cases
the first critical rotational frequency is associated
with a one-dimensional rotation. We use $\Sigma_1<\Sigma_2<\Sigma_3$.
Exploiting that at the transition
point $\omega_y=\omega_z=\omega_{\perp}$, we obtain from Eq.(\ref{cond})
a third order equation for $u=\Omega /\omega _{\perp}$ which reads 
(see also \cite{TA})
\beq
\label{cub}
u^3-\frac{u}{2}+\frac{1}{2}\frac{r-1}{r+1}=0
\eeq
and $r=\Sigma _3 / \Sigma _2$.
From the discriminant of Eq.(\ref{cub}) we obtain
the critical value 
\beq
r_{\it cr}=
\frac{\sqrt{27}+\sqrt{2}}{\sqrt{27}-\sqrt{2}}.
\label{r1}
\eeq   
It was shown in \cite{TA} that a prolate system becomes
eventually oblate with the rotational axis coinciding with the 
symmetry axis, if $r < r_{\it cr}$. 
In general, for $r < r_{\it cr}$, we have 
 the three solutions for Eq.({\ref{cub})
\beq
\label{ph1}
u_1=\sqrt{\frac{2}{3}} \cos \frac{\chi}{3},\quad
u_{2,3}=-\frac{1}{2}\sqrt{\frac{2}{3}}
\left(\cos \frac{\chi}{3} \mp \sqrt{3}\sin \frac{\chi}{3}\right)\\
\eeq
\beq
\label{ang}
\cos \chi= -3\sqrt{\frac{3}{2}}\frac{r-1}{r+1}, \quad 
\pi/2< \chi <\pi .\nonumber
\eeq

 By their derivation these values correspond to three bifurcation
points. 
We recall that for one--dimensional rotation 
the mean field equation Eq.(\ref{cub}) determines
the positive signature solutions. 
At the left and right hand boundary values of the angle 
$\chi$ we obtain from Eqs.(\ref{ph1}), (\ref{ang})
\bea
\chi=\pi/2 \quad (r=1) &\Rightarrow&
\left\{ \begin{array}{ll}
\Omega_1=\omega_\perp/\sqrt{2}\\
\Omega_2=\quad 0\\
\Omega_3=-\omega_\perp/\sqrt{2}\\
\end{array} \right.\\
\chi=\pi \quad (r=r_{\rm cr})&\Rightarrow&
\left\{ \begin{array}{ll}
\Omega_1=\sqrt{2/3}\omega_\perp/2\\
\Omega_2=\sqrt{2/3}\omega_\perp/2 \quad   \\
\Omega_3=-\sqrt{2/3}\omega_\perp \quad\\
\end{array} \right.
\eea
The natural question arises: what is the meaning of the
negative rotational frequencies?
Below we demonstrate that the second solution  
gives the critical point $\Omega_{\it cr}^{(1)}$
at which the lowest positive signature 
vibrational frequency (in the rotating frame) tends to zero.

It should be noted that the solutions $\Omega_{2,3}(u_{2,3})$ 
have been found in \cite{TA},
however, an interpretation was not provided due to
the lack of the RPA analysis.
In \cite{Ak} and \cite{Kur} vibrational excitations  
for one-dimensional rotations around the $x$-axis have been dealt
with by the RPA. However,
in \cite{Ak} the self-consistent condition and the residual interaction
are different from those of the present paper, and
in \cite{Kur} the RPA analysis
was done in the {\it laboratory} frame. We emphasise that 
it is important to treat the RPA equations in the {\it rotating} frame,
since only then the relation between the bifurcation points of the mean 
field and the RPA solutions can be clearly worked out.
This relation has not been analysed in the quoted papers.

\subsubsection{Positive signature solutions}

 As noted above, the determinant $F(\omega_{\lambda})$ 
factors into positive and
negative signature blocks for the oblate rotation.
With the projection $\lambda$ on the symmetry axis $x$
of the angular momentum $L_x$  
being a good quantum number the phonon states with different 
$\lambda$ decouple. Therefore
the secular equation for the positive signature phonons
separates into three independent equations 
each determining the solution for the collective excitations
with $\lambda=0$ and $\lambda=\pm2$.
We focus our analysis on the positive signature excitations
with $\lambda = \pm 2$, since they will define the critical
rotational frequency.

Taking into account the self-consistent condition (\ref{cond})
the secular equation can, after some tedious algebra, be 
presented for $\lambda=+2$ in the form
\beq
\label{s1}
\frac{a}{2+x}+\frac{b}{2-x}+\frac{c}{x}=1 
\eeq
using the notation
\beq
x=\frac{\omega_{\lambda}+2\Omega}{\omega_\perp}, \quad
a=\frac{1}{r+1},\quad b=\frac{r}{r+1},\quad c=\frac{r-1}{r+1}.\nonumber
\eeq
The RPA eigen-frequencies $\omega_{\lambda}$  in the rotating frame 
are obtained by solving the cubic equation 
\beq
x^3-2x+4c=0
\label{pos1}
\eeq
leading to the solution  
\bea
&&\omega_{\lambda=+2}^{(1)}=2\omega_\perp \sqrt{\frac{2}{3}}\cos\frac{\chi}{3}
-2\Omega\\
&&\omega_{\lambda=+2}^{(2,3)}=-\omega_\perp\sqrt{\frac{2}{3}}
\left(\cos \frac{\chi}{3} \mp \sqrt{3}\sin \frac{\chi}{3}\right)
-2\Omega.
\eea
We consider only the first solution as it is the one with
a positive norm. 
Setting this frequency equal to zero we find just
the first bifurcation point 
$u_1=\Omega_1/\omega_\perp$ of Eq.(\ref{cub}) 
\beq
\Omega_1=
\omega_\perp \sqrt{\frac{2}{3}}\cos\frac{\chi}{3}
=\frac{{\tilde \omega}_{\lambda=+2}^{(1)}}{\lambda}
\eeq
where
\beq
{\tilde \omega}_{\lambda=+2}^{(1)}=2\omega_\perp 
\sqrt{\frac{2}{3}}\cos\frac{\chi}{3}.
\label{sp1}
\eeq 

For $\chi =\pi/2 \;(r=\Sigma_3/\Sigma_2=1$)
we obtain from Eq.(\ref{sp1}) the well known estimate of the 
isoscalar giant quadrupole resonance energy
in non-rotating axially deformed nuclei, 
i.e.~$\tilde{\omega}_{\lambda=+2}^{(1)}=\sqrt{2}\omega_\perp$ 
\cite{SR} with 
$\omega_\perp$ being the confinement frequency perpendicular to 
the symmetry axis of the mean field potential. 
The value $\chi=\pi$ yields
$\tilde{\omega}_{\lambda=+2}^{(1)}=\sqrt{2/3}\omega_\perp$.
For a fixed configuration $\Sigma_1<\Sigma_2<\Sigma_3$ 
the oblate state has the maximal
angular momentum $L_x=\Sigma_3-\Sigma_2$ 
(ground-band termination state) \cite{NR}.
If the signature symmetry is conserved
the mode $\omega_{\lambda=+2}^{(1)}$ can make 
a transition to a different configuration with two more
units of angular momentum than that
of the oblate (yrast) state, assuming that the rotational
frequency exceeds the critical frequency $\Omega_1$. 
Therefore, for a {\it fixed} configuration 
the first bifurcation point is related to the quadrupole 
vibrational mode ${\tilde \omega}_{\lambda=+2}^{(1)}$ in the 
laboratory frame carrying two units ($\lambda=+2$) of 
angular momentum.

For the $\lambda=-2$ excitations the secular equation 
is similar to that of Eq.(\ref{s1}) with $x$ replaced by $-y$ where
\beq
y=\frac{\omega_{\lambda}-2\Omega}{\omega_\perp}.
\eeq

The RPA eigen-frequencies $\omega_{\lambda}$ are obtained 
by solving the cubic equation for $y$ 
\beq
y^3-2y-4c=0
\label{pos2}
\eeq
leading this time to the solutions  
\bea
&&\omega_{\lambda=-2}^{(1,3)}=2\Omega+\omega_\perp\sqrt{\frac{2}{3}}
\left(\cos \frac{\chi}{3} \pm \sqrt{3}\sin \frac{\chi}{3}\right)
\\
&&\omega_{\lambda=-2}^{(2)}= 2\Omega-2\omega_\perp \sqrt{\frac{2}{3}}
\cos\frac{\chi}{3}.
\eea

The positive norm solutions are the first and the third.
Setting $\omega_{\lambda=-2}^{(1,3)}$ equal to zero, we obtain
the second and third bifurcation points
$u_{2,3}=\Omega_{2,3}/\omega_\perp$ of Eq.(\ref{cub}) which read
\bea
&&\Omega_2=
-\frac{\omega_\perp}{2}\sqrt{\frac{2}{3}}
\left(\cos \frac{\chi}{3}- \sqrt{3}\sin \frac{\chi}{3}\right)
= \frac{{\tilde \omega}_{\lambda=-2}^{(3)}}{\lambda}\\
&&\Omega_3=
-\frac{\omega_\perp}{2}\sqrt{\frac{2}{3}}
\left(\cos \frac{\chi}{3}+ \sqrt{3}\sin \frac{\chi}{3}\right)
= \frac{{\tilde \omega}_{\lambda=-2}^{(1)}}{\lambda}\\
&&{\tilde \omega}_{\lambda=-2}^{(1,3)}=
\omega_\perp\sqrt{\frac{2}{3}}
\left(\cos \frac{\chi}{3}\pm \sqrt{3}\sin \frac{\chi}{3}\right).
\eea
The frequencies 
${\tilde \omega}_{\lambda=-2}^{(1,3)}$ are the vibrational
modes in the laboratory 
frame carrying $\lambda=-2$ units of angular momentum.

A collective vibration with the projection of $K\neq0$ units of 
angular momentum along the symmetry axis may be associated with 
a surface wave travelling around this axis.
This wave can travel in opposite direction
(negative rotational frequency) with respect to  the
rotation of the core (the vacuum) \cite{Mar96}.

With $\chi=\pi/2\;(\Sigma_3=\Sigma_2)$ we
obtain ${\tilde \omega}_{\lambda=-2}^{(1)}=\sqrt{2}\omega_\perp
={\tilde \omega}_{\lambda=+2}^{(1)}$ 
and ${\tilde \omega}_{\lambda=-2}^{(3)}=0$. 
In this case the angular momentum of the vacuum (oblate) 
state is zero, since $L_x=\Sigma_3-\Sigma_2=0$.
Thus, in the oblate system with zero angular momentum  
a breaking of the symmetry arises at the
positive/negative rotational frequency

\beq
\Omega_{\rm cr}=\tilde{\omega}_{\lambda=\pm2}^{(1)}/\lambda=
\pm\omega_\perp/\sqrt{2}.
\label{axb}
\eeq
It is at this point where the transition from non-collective rotation
(around $x$-axis) to triaxial collective rotation takes place, where the
latter state has two units of angular momentum.

At the other end of the parameter range, $\chi=\pi\;(r=r_{cr})$, we 
obtain ${\tilde \omega}_{\lambda=-2}^{(1)}=2\sqrt{2/3}\omega_\perp$
 and ${\tilde \omega}_{\lambda=-2}^{(3)}=-\sqrt{2/3}\omega_\perp$.
Within the full range of parameter values 
$\pi/2 < \chi <\pi\;(1<r<r_{\rm cr})$
the vibrational modes
$\tilde{\omega}_{\lambda=-2}^{(1,3)}$ are sweeping over the values
\bea
\sqrt{2}\omega_{\perp} & <\tilde{\omega}_{\lambda=-2}^{(1)} < &
2\sqrt{\frac{2}{3}}\omega_\perp,\\
-\sqrt{\frac{2}{3}}\omega_\perp & <\tilde{\omega}_{\lambda=-2}^{(3)} < & 
\quad 0.
\eea

When the rotational frequency approaches 
 the lowest critical value for a triaxial nucleus which is
\beq
\Omega_2 = -\frac{\omega_\perp}{2}\sqrt{\frac{2}{3}}
\left(\cos \frac{\chi}{3}- \sqrt{3}\sin \frac{\chi}{3}\right)
= \Omega _{{\rm crit}}^{(1)} > 0,
\eeq
the vibrational excitation in the rotating
frame for the oblate state tends to zero, i.e., 
$\omega_{\lambda=-2}^{(3)}=\tilde{\omega}_{\lambda=-2}^{(3)}+
2\Omega|_{\Omega=\Omega_2}=
-|\tilde{\omega}_{\lambda=-2}^{(3)}|+2\Omega_{\rm crit}^{(1)}=0$.
The {\it negative} quadrupole mode energy 
$\omega _{\lambda=-2}^{(3)}$ vanishes at
the transition from the non-collective to the
collective triaxial  rotation and becomes positive for the oblate
state at higher rotational frequencies of the mean field.
In other words, the mode $\omega _{\lambda=-2}^{(3)}$ can be viewed 
as a de-excitation in the intrinsic frame
of the vacuum (oblate) state thus leading to the triaxial state with two units
of angular momentum less than the vacuum state. 
In contrast, the mode $\omega_{\lambda=-2}^{(1)}$ is positive
at $\Omega _{{\rm crit}}^{(1)}$ and
describes an ordinary quadrupole excitation with
two units of angular momentum less than 
that of the yrast state on which it is based.

\subsubsection{Negative signature solutions}
 
From the previous subsection we have 
learned that the critical rotational frequency is
related to the vanishing of the RPA solutions in the intrinsic frame
and consequently to the breaking of the mean field symmetry. 
Since the negative signature solutions are associated 
with the $\lambda=\pm1$ excitations, at least one of 
the solutions must be related to signature symmetry breaking, 
i.e.~to the bifurcation point where tilted rotation is setting in.

In the oblate case the secular equation for the negative signature
excitations contains the "spurious" solution at $\omega=\Omega$
corresponding to a collective rotation and the vibrational modes; 
the latter include the wobbling motion and 
carry one unit of angular momentum \cite{JM,Mar}.
Taking into account the self-consistent condition (\ref{cond})
the secular equation for the $\lambda=-1$ excitation reads
\bea
\label{s3}
&&\frac{e}{\omega_x+\omega_\perp+\Omega-x}+
\frac{f}{\omega_x+\omega_\perp-\Omega +x}\nonumber\\
&&+\frac{d}{\omega_x-\omega_\perp-\Omega +x}
+\frac{p}{\omega_x-\omega_\perp+\Omega -x}= \frac{2}{\omega_x} 
\eea
where we use the notation
\bea
e&=&\frac{1+k}{r+1}, \quad
f=\frac{r+k}{r+1},\nonumber \\
d&=&\frac{1-k}{r+1},
\quad p=\frac{r-k}{r+1}, \quad k=\Sigma_1/\Sigma_2. \nonumber
\eea
The root in $x$ of Eq.(\ref{s3}) is the RPA frequency 
$\omega_\lambda$.
After some simplification we obtain 
\beq
y(y^3-y-2\alpha c)=0,
\label{neg1}
\eeq
where 
\beq
y=\frac{\omega_\lambda-\Omega}{\sqrt{\omega_x^2+\omega_\perp^2}},\quad
\alpha=\frac{\omega_x^2\omega_\perp}{(\omega_x^2+\omega_\perp^2)^{3/2}}.
\eeq

The trivial solution $y=0$ corresponds to the
"spurious" excitation $\tilde{\omega}_{\lambda}=0$
in the laboratory frame with
the mode $ {\omega}_{\lambda}=\tilde{\omega}_{\lambda}+\Omega$
being a rotational mode 
with frequency $\Omega$ in the rotating frame.
For Eq.(\ref{neg1}) there exist three other solutions  
\bea
&&\omega _{n}=2 \sqrt{ \frac{\omega_x^2+\omega_\perp^2}{3}} 
\cos {\frac{\psi +2\pi n}{3}}+\Omega,\quad n=0,1,2
\label{ns1}\\
&&\cos \psi =
\sqrt{27}\frac{\omega_x^2\omega_\perp }{(\omega_x^2 +\omega_\perp^2)^{3/2}}
\frac{r-1}{r+1},\quad 0< \psi < \pi/2
\label{psi}
\eea
The first solution ($n=0$) is the one with the positive norm.
Setting it equal to zero, we obtain the first bifurcation 
point for the negative signature
\beq
\Omega_1=-2 \sqrt{ \frac{\omega_x^2+\omega_\perp^2}{3}} 
\cos {\frac{\psi}{3}}=\frac{\tilde{\omega}_{\lambda=-1}^{(1)}}{\lambda}.
\eeq
Here, the mode
\beq
{\tilde{\omega}}_{\lambda=-1}^{(1)}= 
2 \sqrt{ \frac{\omega_x^2+\omega_\perp^2}{3}} 
\cos {\frac{\psi}{3}}
\eeq
describes the quadrupole excitation in the laboratory frame,
which is based on the yrast (oblate) state 
and carries one unit of
angular momentum, i.e.~$\lambda=-1$.
Since $\cos {\frac{\psi}{3}}> \sqrt{3}/2$, $\Omega_1$ is larger than
the maximal rotational frequency (see Appendix B), it cannot
be associated with a stable tilted rotation.
For $\Sigma_3=\Sigma_2 \; (\psi=\pi/2)$
the mode $\tilde{\omega}_{\lambda=-1}^{(1)}=\sqrt{ \omega_x^2+\omega_\perp^2}$
is again the isoscalar quadrupole giant resonance energy 
in non-rotating axially deformed nuclei \cite{SR}. 
Notice that this mode has been introduced in the context of a 
BEC under the name "scissor mode" \cite{GS}.
In the spherical case it is
$\tilde{\omega}_{\lambda=-1}^{(1)}=\sqrt{2}\omega_0$ \cite{BM75}. 
With $\psi=0$ it is 
$\tilde{\omega}_{\lambda=-1}^{(1)}=2\sqrt{(\omega_x^2+\omega_\perp^2)/3}$.   
   
For the $\lambda=1$ excitations the secular equation
is similar
to Eq.(\ref{s3}) with $x$  replaced by $-x$.
After simplification, we obtain
\beq
y(y^3-y+2\alpha c)=0, \quad y=
\frac{\omega_\lambda+\Omega}{\sqrt{\omega_x^2+\omega_\perp^2}}.
\label{neg2}
\eeq
Again, the trivial solution $y=0$ corresponds to the
"spurious" excitation $\tilde{\omega}_{\lambda}=0$
in the laboratory frame with 
the mode $ {\omega}_{\lambda}=\tilde{\omega}_{\lambda}-\Omega$
being a rotational mode 
with frequency $-\Omega$ in the rotating frame.
The three other solutions of Eq.(\ref{neg2}) are
\bea
&&\omega_{\lambda=+1}^{(1,2)}=\sqrt{\frac{\omega_x^2+\omega_{\perp }^2}{3}}
\left(\cos \frac{\psi}{3} \mp \sqrt{3}\sin \frac{\psi}{3}\right)-\Omega
\label{mod1}
\\
&&\omega_{\lambda=+1}^{(3)}=-2\sqrt{\frac{\omega_x^2+\omega_{\perp }^2}{3}}
\cos\frac{\psi}{3}-\Omega.
\eea

The positive norm solutions are the first and the second.
Setting $\omega_{\lambda=+1}^{(1,2)}$ equal to zero, we obtain the 
bifurcation points
\bea
&&\Omega_2=
\sqrt{\frac{\omega_x^2+\omega_{\perp }^2}{3}}
\left(\cos \frac{\psi}{3}- \sqrt{3}\sin \frac{\psi}{3}\right)
= \frac{{\tilde \omega}_{\lambda=+1}^{(1)}}{\lambda}\\
&&\Omega_3=
\sqrt{\frac{\omega_x^2+\omega_{\perp }^2}{3}}
\left(\cos \frac{\psi}{3}+ \sqrt{3}\sin \frac{\psi}{3}\right)
= \frac{{\tilde \omega}_{\lambda=+1}^{(2)}}{\lambda}\\
&&{\tilde \omega}_{\lambda=+1}^{(1,2)}=
\sqrt{\frac{\omega_x^2+\omega_{\perp }^2}{3}}
\left(\cos \frac{\psi}{3}\mp \sqrt{3}\sin \frac{\psi}{3}\right).
\eea
Thus, they are related to the vibrational 
modes ${\tilde \omega}_{\lambda=+1}^{(1,2)}$ in the laboratory 
frame and carry $\lambda=+1$ units of angular momentum.

In fact, the lowest 
mode $\omega_{\lambda=+1}^{(1)}(\Omega)$ corresponds to a wobbling 
excitation in the intrinsic frame.
As soon as the rotational
frequency approaches the lowest value $\Omega_2$,
the mode $\omega_{\lambda=+1}^{(1)}= 
{\tilde \omega}_{\lambda=+1}^{(1)}-\Omega$ 
becomes soft and signals the onset of genuine 
tilted rotation of the nucleus. 
The mode energy vanishes for
\begin{equation}
\Omega_2
= 2\sqrt{\frac{\omega
_x^2+\omega_{\perp }^2}{3}}\cos {\psi +\pi \over 3}=
\Omega _{{\rm crit}}^{(2)}.
\label{cr2}
\end{equation}
Thus, the bifurcation frequency $\Omega_2$ corresponds to the 
transition from  dynamical fluctuations of the angular momentum
in the oblate system 
to the stable tilted rotation of the triaxial shape.
This critical frequency coincides perfectly with our numerical
results.

\begin{figure}
\epsfxsize=2.2in
\centerline{
\epsffile{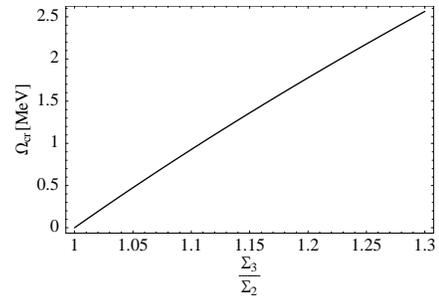}}
\vglue 0.15cm
\caption{ Critical frequency $\Omega _{{\rm crit}}^{(2)}$ versus 
$r=\Sigma _3/\Sigma _2$
for a typical (near) oblate situation. The example refers to $N=16$ with 
$(\Sigma _1,
\Sigma _2,\Sigma _3)=(30,40,44)$; with $\omega _x=20.7,
\omega _{\perp}=14.5$ it is
$\Omega _{{\rm crit}}^{(2)}=0.93$ for $r=1.1$.
}
\label{fig2}
\end{figure}

Furthermore, Eq.(\ref{cr2}) explains why oblate nuclei
(at $\Omega=0$) cannot make
the transition to tilted  rotation. In fact, for these nuclei
$r=1$ and hence $\psi =\pi /2$ in Eq.(\ref{psi}). 
As a
consequence, $\Omega _{{\rm crit}}^{(2)}$ is zero 
($\tilde{\omega}_{\lambda=+1}^{(1)}=0$), 
i.e.~no transition can occur with our convention
$\Sigma_1<\Sigma_2=\Sigma_3$. In contrast, whenever $r>1$, a positive value is
obtained for $\Omega _{{\rm crit}}^{(2)}$ as illustrated in Fig.2.

With $r=1$, i.e.~$L_x=\Sigma_3-\Sigma_2=0$, 
the other two modes $\lambda=\pm1$ are 
degenerate, since
$\tilde{\omega}_{\lambda=+1}^{(2)}=\sqrt{\omega_x^2+\omega_\perp^2}=
\tilde{\omega}_{\lambda=-1}^{(1)}$. 
The corresponding critical frequencies
\beq
\Omega_{\rm cr}=\tilde{\omega}_{\lambda=\pm 1}/\lambda =
\pm\sqrt{\omega_x^2+\omega_\perp^2}
\label{obb}
\eeq
are larger than the maximal frequency 
(see Appendix B) and do not 
give rise to symmetry breaking of the mean field.
Nevertheless, these critical 
frequencies can be related to vibrational excitations as
discussed in the following.

As was suggested by Marshalek and Sabato \cite{MS} 
the self-consistent cranking model should 
allow the study of {\it vibrational}
states in spherical nuclei rotating around the symmetry axis.
Our results for quadrupole excitations demonstrate that 
this idea can be succesfully employed for any axially deformed 
system rotating around the symmetry axis (see also a 
general discussion in \cite{Mar96}).
The vibrational excitations with
nonzero angular momentum $\lambda$ (or $K$) 
aligned along the symmetry axis
are determined by the bifurcation points 
of the mean field via Eq.(\ref{mb})
\beq
\tilde{\omega}_{\lambda}=\lambda \Omega_{\rm cr}.
\eeq
The lowest critical rotational frequency 
(bifurcation point) gives the
collective excitation with the
largest angular momentum along
the symmetry axis.
If the critical rotational frequency 
lies below the maximal rotational frequency,
it will give rise to a
mean field symmetry breaking, as the corresponding 
vibrational excitation vanishes in the rotating frame.
If the frequencies lie above the maximal
value, they yield the higher vibrational excitation
energies with nonzero angular momentum in the laboratory frame. 
They persist as their counterparts in the
rotating frame do not vanish.

We have
established numerically in all cases of triaxial oblate nuclei
the onset of tilted rotation and identified exactly the value of
Eq.(\ref{cr2}) as the lowest rotational frequency. The
values of $\Omega _{{\rm crit}}^{(2)}$ are all fairly small in
these cases ($\Omega _{{\rm crit}}^{(2)}<1$), since the values for
$r$ are close to unity. 
This is in contrast to light prolate nuclei to which we turn next.
For prolate nuclei (at $\Omega=0$) we have $\Sigma_1=\Sigma_2
\neq \Sigma_3$ and the oscillator frequencies are $
\omega_x^0=\omega _y^0= \omega _{\perp}^0$ implying the
symmetry axis to be the $z$-axis. Since in this case $r>1$, Fig.2 cannot serve
as a basis to exclude tilted rotation. Yet we present an
independent argument.

\subsection{Critical frequencies for prolate rotation}

Consider a rotation around
the symmetry axis, i.e.~the $z$--axis. The eigenmodes have a
particular simple form \cite{Val,RBK,Zel} and Eq.(\ref{cond}) leads to
the nontrivial solution which must satisfy the equation 

\beq
(\omega_ + \omega_- -\Omega^2)=0. 
\eeq 

Setting
$\omega_x=\omega_y=\omega _{\perp}^0$ we obtain the bifurcation
point \cite{Mar93} 

\beq 
\label{b1} 
\Omega_{{\rm cr}}=\frac{\omega_{\perp}^0}{\sqrt{2}}. 
\eeq 

For $\Omega >\Omega_{{\rm cr}}$ the axial symmetry is broken 
and the system is driven into the domain of triaxial shape 
under the PAC rotation.
This situation is reminiscent of a striking feature established
experimentally in the rotating BEC \cite{4,5,6}; there the
minimal rotational frequency for nucleation of a vortex occurs at
around $0.7 \omega_{\perp}$, where  $\omega_{\perp}$ is the mean
oscillation frequency of atoms in the $x-y$ plane, irrespective
of the number of atoms or oscillation frequency $\omega_z$ along
the $z$--axis. According to our analysis, the dynamical instability
in a nuclear system and in the rotating BEC is of a similar
nature, in spite of the different character of the interaction
between nucleons (attractive) and between atoms (repulsive). The
important aspect is the trapping of either system by the harmonic
oscillator potential. For the nuclear system it is the mean field
and for the BEC it is the external magnetic field.

For a prolate system the projection $K$ of angular momentum is a
good quantum number as $[L_z,O_K^{\dagger}]=KO_K^{\dagger}$. 
We thus obtain an equation similar to Eq.(\ref{man}) after
replacing $\lambda$ by $K$.
At the rotational frequency $\Omega_{\rm
crit}={\tilde \omega}_K/K$ one of the RPA frequency vanishes. 
For a prolate rotation the RPA equations are solved separately
for each quadrupole mode, since there is no coupling between
different $K$ modes. 

For the quadrupole phonons with the largest projection $K=2$ it is
\beq 
\label{K2}
\omega_{K=2}=\sqrt{2}\omega_{\perp}-2\Omega.
\eeq
This mode is the quadrupole excitation having two more units of
angular momentum than the vacuum state. When $\omega_{K=2}=0$ the
transition from non-collective rotation (around the $z$--axis) to
triaxial collective rotation takes place, i.e.~at 
\beq
\Omega_{{\rm cr}}={\omega_{\perp}^0\over \sqrt{2}} 
\label{prb}
\eeq 
which is just the bifurcation point of the mean field in Eq.(\ref{b1}).
This bifurcation
point applies for any axially symmetric system (prolate and oblate) as
follows from Eqs.(\ref{axb}) and (\ref{prb}).

For the $K=1$ mode the RPA solution is

\beq 
\label{K1}
\omega_{K=1}=
\sqrt{\omega_{\perp}^2+\omega_z^2}-\Omega.
\eeq

The condition $\omega_{K=1}=0$ yields the critical frequency at
which the onset of the tilted rotation should occur for the
prolate system, i.e.~at 

\beq 
\label{tz} 
\Omega_{{\rm cr}}=\sqrt{\omega_{\perp}^2+\omega_z^2}. 
\eeq 

However, the
energy to create this mode of angular momentum $1\hbar $ is too high,
the system rather prefers the PAC rotation around the axis
perpendicular to the symmetry axis. Besides, 
as discused above, the value of the
rotational frequency in Eq.(\ref{tz}) lies outside the physical
range of rotational frequencies (see Appendix B).
As a consequence,
our model does not allow tilted rotations for systems which are
prolate at $\Omega=0$, i.e.~if $\Sigma_1=\Sigma_2$. 
In contrast to the rotational frequency defined by 
Eqs.(\ref{axb}), (\ref{prb}), 
this frequency is related to vibrational excitations 
carrying one unit of angular momentum in 
an axially symmetric system,
be it oblate (Eq.(\ref{obb})) or prolate (Eq.(\ref{tz}))}.

Note that the assumption $\Sigma_1=\Sigma_2$ is essential for the
arguments presented here. 
In contrast, if all three values of 
$\Sigma_i$ differ from each other, tilted rotations do occur at the
value given by Eq.(\ref{cr2}) just as it has been found for the near
oblate situations. We recall, however, that this happens for
$N<56$ at values so high that the corresponding energy minimum is
only local. In contrast, when $N>70$, the relevant value of
$\Omega _{{\rm crit}}^{(2)}$ is sufficiently small so as yield a
global minimum.

\section{Summary}

Within the model considered only nuclei being triaxial in their ground state
exhibit tilted rotation at a specific rotational frequency. In fact,
the analytic expressions for these critical frequencies
 obtained in the present paper are
in a perfect agreement with our numerical findings.
 All nuclei attain at $\Omega _{{\rm crit}}^{(1)}$ an oblate shape.
The value of $\Omega _{{\rm crit}}^{(1)}$ depends on the particular
nucleus. For light and medium prolate or near prolate nuclei $(N<56)$ this energy
minimum is a local minimum, while for oblate or near oblate nuclei and heavy
(near) prolate $(N>70)$ it is a
global minimum. All triaxial nuclei -- these are the nuclei which form
the little `bubbles' in Fig.1 -- go through a second transition and
admit a tilted rotation beyond $\Omega _{{\rm crit}}^{(2)}$ while the
tilted rotation cannot occur when starting with a
nucleus that is axially symmetric at $\Omega =0$. 
From our model, the occurrence and identification of a
tilted rotation must be seen as a signature for triaxiality.
It is true that the harmonic oscillator model is a simplistic approximation
for the phenomenon
of the tilted rotation. However, our findings may provide guidance
for values of particle number and rotational frequency at which
tilted rotations may be found.

Using the cranking approach to describe non-collective rotation around 
a symmetry axis we have established the
correspondence between symmetry-breaking bifurcation points of the mean
field and the RPA frequencies carrying non-vanishing angular momentum aligned
along the symmetry axis. We have demonstrated that the dynamical instability
in a nuclear system and in the rotating BEC are of a similar nature,
in spite of the different characters of the interaction
between nucleons (attractive) and between atoms (repulsive).
The remarkable agreement with regard to the bifurcation points
in both systems is due to the common nature of the confining potential
and its dependence on the rotational speed.

\vskip 2cm

R.G.N. acknowledges financial support from the National
Research Foundation of South Africa which was provided under the
auspices of the Russian/South African Agreement on Science and
Technology. He is also thankful for the warm hospitality which he
received from the Department of Physics during his visit
to South Africa. This project has been supported in part
by the RFBR under the Grant 00-02-17194.

\appendix
\def\theequation{\thesection.\arabic{equation}}
\section{Eigenmodes}
\setcounter{equation}{0}
The Hamilton function of Eq.(\ref{ham}) can be written in matrix form
\begin{equation}
H=\{ \vec p,\vec r \}^T{\cal H} \{ \vec p,\vec r \}
\end{equation}
where $\vec p$ and $\vec r$ are combined to the six dimensional column
vector $\{ \vec p,\vec r\}$ and
\begin{equation}
{\cal H}=\pmatrix{1&0&0&0&-\Omega _z&\Omega _y \cr
0&1&0&\Omega _z&0&-\Omega _x \cr
0&0&1&-\Omega _y&\Omega _x&0 \cr
0&\Omega _z&-\Omega _y&
\omega _x^2&0&0\cr
-\Omega _z&0&\Omega _x&0&
\omega _y^2&0\cr
\Omega _y&-\Omega _x&0&0&0&
\omega _z^2 }.
 \end{equation}
We aim at the quantum mechanical form in terms of boson operators
\begin{equation}
H= \{Q,Q^{\dagger }\}^T {\cal H}_{qm} \{Q,Q^{\dagger }\}=
\sum E_j(Q^{\dagger }_jQ_j+1/2)
\label{bos}
\end{equation}
where we denote by $\{Q,Q^{\dagger }\}$ a column vector which
is the transpose of the vector
$(Q_1,Q_2,Q_3,Q_3^{\dagger },Q_2^{\dagger },Q_1^{\dagger })$ and
where
\begin{equation}
{\cal H}_{qm}={1\over 2}\pmatrix{0&0&0&0&0&E _1 \cr
0&0&0&0&E _2&0 \cr
0&0&0&E _3&0&0 \cr
0&0&E _3&0&0&0 \cr
0&E _2&0&0&0&0 \cr
E _1&0&0&0&0&0 }.
\label{skew}
\end{equation}

It is instructive to obtain the eigenmode energies from a purely classical
calculation. This approach also provides the linear transformation
between the $Q_j$ and the $\{p_i,x_i\}$.

The classical equations of motion
for the Cartesian components of the momentum and position coordinates read
\begin{equation}
{d\over dt}\pmatrix{{\vec p} \cr {\vec r}}= {\cal M}
\pmatrix{{\vec p}\cr {\vec r}}
\label{eqm1}
\end{equation}
 The matrix ${\cal M}$ is given by
\begin{equation}
{\cal M}=\pmatrix {0&-I_3\cr I_3&0}{\cal H}  \label{calm}
\end{equation}
where $I_3$ is a 3 by 3 unit matrix. The classical orbits are
the solution of Eq.(\ref{eqm1}) which reads
\begin{equation}
\{ \vec p(t),\vec r(t)\} =
{\cal U}\exp({\cal D}t){\cal V} \{\vec p(0),\vec r(0)\}
\end{equation}
where ${\cal D}={\rm diag}(-iE_1,-iE_2,-iE_3,iE_3,iE_2,iE_1)$ is the
diagonal form of $\cal M$. The initial conditions
$\{\vec p(0),\vec r(0)\}$ are of no interest here.
From Eq.(\ref{calm}) it follows that the eigenmodes listed in
Eq.(\ref{skew}) occur in the classical problem in this form.
They are obtained from the secular equation
\begin{equation}
\det|E I_6-{\cal M}|=0.
\end{equation}
The above equation turns out to be a third order polynomial in $E ^2$
and has also been found by \cite{VH}. The column
vectors of $\cal U$ are the (complex) right hand eigenvectors of $\cal M$.
Note that $\cal M$ is not symmetric, hence neither $\cal U$ nor $\cal V$
are unitary, yet ${\cal V}= {\cal U}^{-1}$. We denote
the column vectors of $\cal U$ by $u^{(k)}$, they obey the equations
\begin{eqnarray}
({\cal M}+iE _kI)u^{(k)}&=&0,\quad k=1,2,3 \nonumber \\
({\cal M}-iE _{7-k}I)u^{(k)}&=&0,\quad k=4,5,6
\label{eigvec} \end{eqnarray}
which can be solved as an inhomogeneous system by choosing an arbitrary
component of $u^{(k)}$ equal to unity. The proper normalisation is achieved
by the observation that, up to normalisation
factors, the matrix ${\cal V}={\cal U}^{-1}$ can be
written as
\begin{equation}
{\cal V}=\pmatrix{0&0&0&0&0&-i \cr
0&0&0&0&-i&0 \cr
0&0&0&-i&0&0 \cr
0&0&i&0&0&0 \cr
0&i&0&0&0&0 \cr
i&0&0&0&0&0 }{\cal U}^T\pmatrix {0&-I_3\cr I_3&0}.
\end{equation}
This implies that ${\cal U}^T \cal H U$ is in fact skew-diagonal as
in Eq.(\ref{skew}), and therefore $\cal U$ can be normalised such that
${\cal U}^T {\cal H U}={\cal H}_{qm}$. This implies in particular
\begin{equation}
\{\vec p,\vec r\}={\cal U} \{Q,Q^{\dagger }\}.  \label{lin}
\label{trans}
\end{equation}

The properties of the matrix ${\cal U}$ guarantee the commutator
$[Q_k,Q_{k'}^{\dagger }]=\delta _{k,k'}$ as a consequence of
$[x_i,p_j]=i\hbar \,\delta _{i,j}$. Clearly, for $\Omega =0$ the $Q_j$
become the usual operators
$a_i=(p_i-im\omega _ix_i)/\sqrt{2m \hbar \omega _i}$.

\section{Maximal rotational frequency}
\setcounter{equation}{0}
At particular values of $\Omega$ the roots of cubic equation become negative,
i.e., the corresponding eigenmodes $E_k$ would be imaginary.
Let us consider the simple case of a rotation about the principal axis $x$.
The solutions for the eigenmodes are well known \cite{Val,RBK,Zel}
\beq
E_1=\omega_x; \quad E_2=\omega_+; \quad E_3= \omega_-
\eeq
\beq
\omega_{\pm}^2=\frac{\omega_y^2+\omega_z^2}{2}+\Omega^2 \pm
\frac{1}{2}\Bigg[(\omega_y^2-\omega_z^2)^2+8\Omega^2(\omega_y^2+\omega_z^2)
\Bigg]^{1/2}
\eeq
Thus, for the eigenmodes to be real, $\omega_y$,
$\omega_z$ and $\Omega$ must satisfy
\beq
\frac{\omega_y^2+\omega_z^2}{2}+\Omega^2\geq
\frac{1}{2}\Bigg[(\omega_y^2-\omega_z^2)^2+
8\Omega^2(\omega_y^2+\omega_z^2)\Bigg]^{1/2}
\eeq
which leads to the inequality
\beq
(\Omega^2-\omega_y^2)(\Omega^2-\omega_z^2)\geq 0.
\eeq
This means that the frequency $\Omega$ must be either larger or
smaller than both frequencies.
Since the condition $\Omega \geq \sup(\omega_y$, $\omega_z)$
is unphysical, the rotational frequency should not exceed the
maximal rotational frequency $\Omega_{{\rm max}}\leq \inf(\omega_y$, $\omega_z)$.

\section{Quadrupole operators}
For a solution of the RPA equation of motion it is convenient
to use the following representation of the quadrupole operators
\bea
&&D_0=\sqrt{\frac{5}{16\pi}}(2\bar z^2-\bar x^2-\bar y^2)\\
&&D_1 = \frac{i}{\sqrt{2}}(D_{21}+D_{2-1})=
\sqrt{\frac{15}{4\pi}}{\bar y}{\bar z}\\
&&D_2 = \frac{1}{\sqrt{2}}(D_{22}+D_{2-2})=
\sqrt{\frac{15}{16\pi}}({\bar x}^2-{\bar y}^2)\\
&&D_3 = \frac{1}{\sqrt{2}}(D_{21}-D_{2-1})=
-\sqrt{\frac{15}{4\pi}}{\bar x}{\bar z}\\
&&D_4 =  \frac{i}{\sqrt{2}}(D_{22}-D_{2-2})=
-\sqrt{\frac{15}{4\pi}}{\bar x}{\bar y}
\eea 
With these definitions we have 
$\sum_{\mu=-2}^2D_{\mu}^{\dagger}D_{\mu}\equiv \sum_{m=0}^4D_m^2$.
Using the transformation Eq.(\ref{trans}) the quadrupole
operators are expressed in terms of the boson operators $Q_i,Q_i^\dagger$.
In case of the one-dimensional rotation the matrix ${\cal U}$ reduces to
the transformation as in \cite{Zel} and all matrix elements are calculated 
analytically. When the signature is a good quantum number, 
the operators can be classified according to their signature 
quantum number: 
the operators $D_0,D_1,D_2$ are characterized by the positive, 
while the operators $D_3,D_4$ are characterized 
by the negative signature quantum number \cite{JM,Mar,KN}.
In the oblate case it is convenient to calculate the matrix
elements using $x$-axis quantization, since the
symmetry axis coincides with the axis of rotation.

\end{document}